\documentclass[prb,twocolumn,floatfix,nobibnotes,superscriptaddress]{revtex4}
\usepackage{amsmath}

\usepackage{graphicx}

\begin{document}


\title{
From collinear to vortex magnetic structures in Mn corrals on Pt(111)
}

\author{M. S. Ribeiro}
\affiliation{Faculdade de F\'\i sica, Universidade
Federal do Par\'a, Bel\'em, PA, Brazil}
\author{G. B. Corr\^ea Jr.}
\affiliation{Faculdade de F\'\i sica, Universidade
Federal do Par\'a, Bel\'em, PA, Brazil}
\author{A. Bergman}
\affiliation{Department of Material Science and Engineering, Royal Institute of Technology, Stockholm, Sweden}
\author{L. Nordstr\"om}
\affiliation{Department of Physics and Astronomy, Uppsala University, Box 516, 75120 Uppsala, Sweden}
\author{O. Eriksson}
\affiliation{Department of Physics and Astronomy, Uppsala University, Box 516, 75120 Uppsala, Sweden}
\author{A. B. Klautau}
\affiliation{Faculdade de F\'\i sica, Universidade
Federal do Par\'a, Bel\'em, PA, Brazil}

\date{\today}

\begin{abstract}

\noindent

We study the magnetic properties of small Mn ring shaped clusters 
on a Pt(111) surface in the framework of density functional theory. 
We find that the Mn atoms possess large magnetic
moments, of the order of 4 $\mu_B$/atom, and have dominating
antiferromagnetic interatomic exchange interactions.  A quantum confinement effect within the ring like clusters was found, indicating that even very small clusters can be seen as quantum corrals.
The antiferromagnetic exchange couplings lead to
collinear magnetic arrangements in simple corrals, as well as complex 
non-collinear ordering, as vortex-like structures, 
for the case of corrals with particular geometry 
where antiferromagnetism becomes frustrated.



\end{abstract}
\maketitle
%
\section{Introduction}

The electronic and magnetic properties of nano-objects are known to be possible to influence and manipulate, by control of the geometry. This has for instance been demonstrated for the electronic properties, by use of Fe corrals on a Cu surface, where the size-induced quantization of the electron states was detected.\cite{eigler} As concerns magnetic properties of nano-objects we note the enhanced anisotropy and magnetic moments of wires of Co on a Pt surface.\cite{gambardella} The possibility to use nano-materials to achieve novel functionality, is clearly one of the driving forces in this field, and one of the focus areas concern small clusters of atoms supported on a substrate. These systems are often characterised by scanning tunneling microscopy (STM)\cite{binnig,bode}, where atomic scale properties are imaged directly.  

One of the excitements of the properties nano-sized objects concerns the possibility to use theoretical models and concepts which are defined in a much more general forum, with applicability to several fields of physics. An example of this is the Skyrmion, which is a characteristic of nonlinear continuum
models with length scales ranging from microscopic to cosmological and which serve as models explaining, e.g., classical liquids and liquid crystals.\cite{skyrmie} In the field of nano-magnetism Skyrmions have been discussed to be driven by chiral interactions.\cite{rossler} Another topological object which is discussed in several fields of physics is the so called Z$_2$ structure or Z$_2$ vortex, which is considered in, e.g., string theory as well as thin film magnetism. As we shall see below, we have identified a cluster with a magnetic structure with this symmetry.

As concerns first principles theory, which is the focus of the present study, we note that previous studies have been made for supported clusters, both as concerns the electronic structure\cite{uzdin,stoeffler} as well as magnetic configurations, where in particular the possibility for non-collinear magnetic structures have been discussed\cite{anders,cluster1,cluster2,cluster3,cluster4}. In this paper we have extended our previous efforts in describing the electronic structure and magnetic properties of supported clusters, to investigate the magnetic 
ordering and interatomic exchange-interactions for corrals of Mn atoms, deposited on a Pt(111) surface. In these studies we have paid special attention to the possibility of forming nano-objects with a magnetic structure of Skyrmion or $Z_{2}$ vortex type, and as we shall see below, we have been able to identify one such system.


\section{Method and Computational Details}
The calculations have been performed using the first-principles, self-consistent  real-space linear muffin-tin orbital 
 method within the atomic sphere approximation (RS-LMTO-ASA)\cite{RS,Imp} that is an  order-\textit{N} method and has 
been extended  to the treatment of non-collinear magnetism\cite{anders}. 
The RS-LMTO-ASA method used  is based on the LMTO-ASA formalism \cite{andersen},  
and uses the recursion method \cite{haydock} to solve the
eigenvalue problem directly in real space.  
A detailed description of the method for
 each situation can be found in the corresponding references
cited.
  All calculations are fully self-consistent
and were performed within the local spin density
approximation (LSDA)\cite{vonbarth}.
Here, we have considered Mn corrals with different geometries
 supported on a Pt(111) surface. The Pt surface was simulated by a cluster of 
$\sim$5000 atoms, where the sites have
been placed on a regular fcc lattice with the experimental lattice
parameter of Pt. A cutoff parameter $LL$=20 was
taken in the recursion chain and a Beer-Pettifor
terminator was used\cite{beer}. 
The vacuum outside the surface was modeled by having 
 two layers of empty spheres above the Pt surface in order to
provide a basis for the wave function in the vacuum and to
treat charge transfers correctly. The calculations of the Mn nanostructures 
have been performed by embedding the clusters as a perturbation
on the  self-consistently converged Pt(111) surface.
The Mn sites and those closest shells of Pt (or empty spheres) 
 sites around the defect were  
recalculated self-consistently, 
 with size varying from 84 up to 102 sites, 
while the electronic structure
for all other sites far from the Mn cluster were kept unchanged.
Structural relaxations have not been included in most of the systems considered in this study.
An example justifying this choice is shown in section III.

In order to calculate the orbital moments, for the collinear solutions, we performed 
fully relativistic calculations where the spin–orbit interaction
is treated at each variational step. 
The method used here can also treat
relativistic effects by including a spin-orbit coupling term to
the Hamiltonian for noncollinear calculations. However, for the systems considered in
this study it was found exceedingly small magnetic anisotropy energies below the numerical uncertainty. 
Therefore, the noncollinear calculations presented here were performed without
including the spin-orbit coupling. 

For a large selection of the considered Mn corrals,  
we have calculated exchange interactions
 directly using the formula of Liechtenstein \textit{et al}.\cite{Lie} 
 as implemented in the RS-LMTO-ASA\cite{soniaJ1}. The exchange coupling parameters 
 (\textit{Jij}'s) shown in this study have
been obtained from the ferromagnetic configuration of the
clusters. These values of the \textit{Jij}'s are used to discuss, 
on a qualitative level, the cause of the magnetic
ordering obtained in the full noncollinear calculation.

\par
\section{Results and discussion}
\begin{figure}[htp]
\begin{center}
\includegraphics*[width=0.9\linewidth]{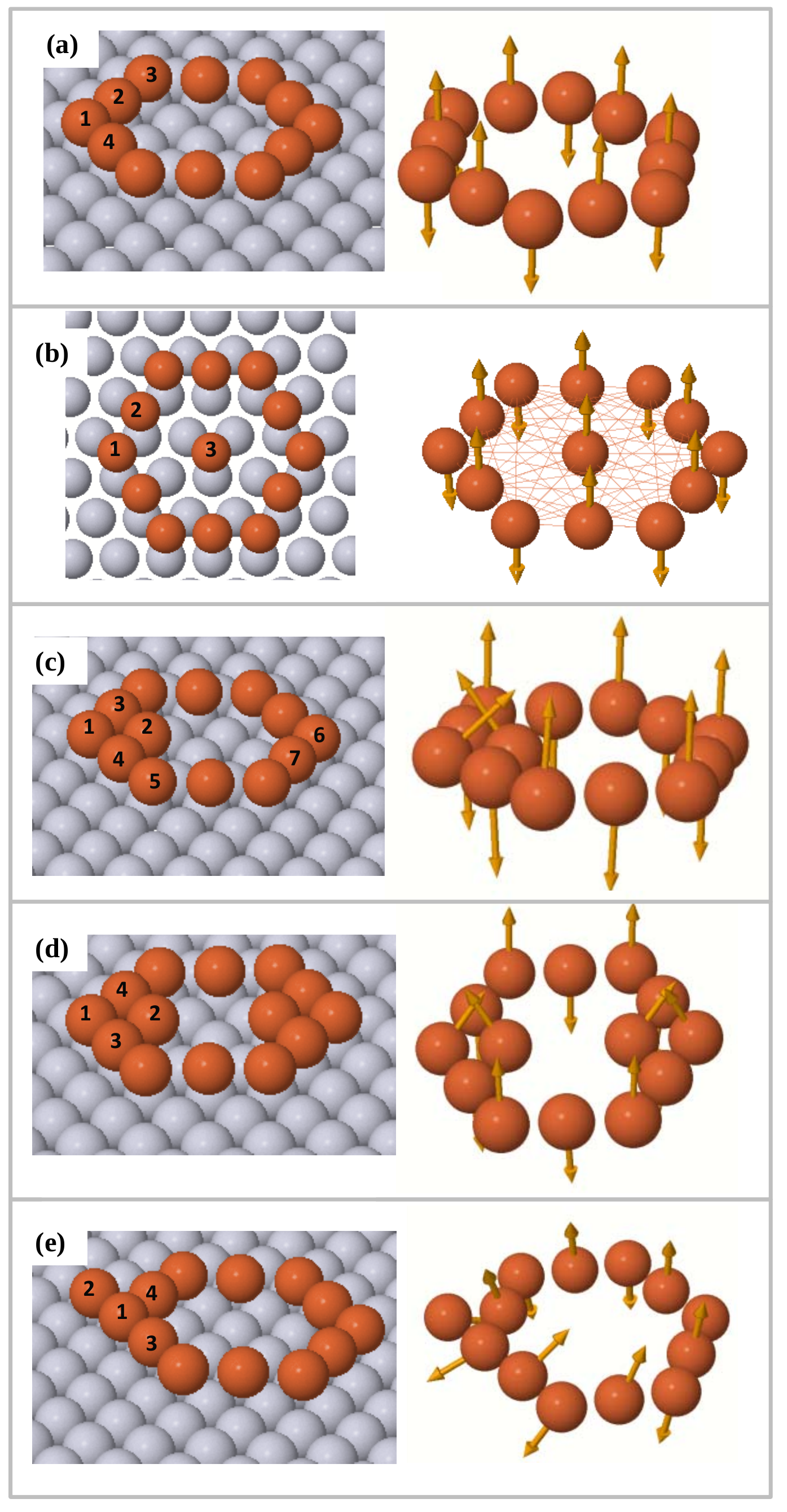}
\caption{\label{corral1abcde}(Color online) The magnetic ordering for Mn corrals on a Pt(111) surface.}
\end{center}
\end{figure}

In Fig.~\ref{corral1abcde}, the structure of several Mn corrals supported on Pt(111) are sketched. The starting point for these geometries is a hexagonal Mn corral composed of 12 Mn adatoms, resulting in a diameter of $d=1.1$ nm, as seen in Fig.~\ref{corral1abcde}(a). This geometry has then been varied by adding one or more additional atoms	 to the structure, Figs.~\ref{corral1abcde}(b-e).
As expected, the magnetic moments for the Mn atoms in the clusters depend on the number of nearest neighbors of Mn, where fewer neighbors yield higher magnetic moment, with values ranging from 4.6$\mu_{B}$/atom, for the central atom at Fig.~\ref{corral1abcde}(b), 
to 4.2$\mu_{B}$/atom, for the Mn atom labeled by 2 on Fig.~\ref{corral1abcde}(c) and (d). 
 For all systems considered in this study, we have found that the orbital   
 moments for the Mn atoms are small, of the order of 0.05$\mu_{B}$ per atom.
 The induced spin magnetic moments at Pt atoms vary from 0.05$\mu_{B}$ to 0.16 $\mu_{B}$, where sites with larger number of Mn nearest neighbors present slightly larger magnetic moments, and the orbital moments at Pt atoms are around 0.02$\mu_{B}$. 

The Mn corral with 12 atoms, shown in Fig.~\ref{corral1abcde}(a), has an antiferromagnetic collinear ground state, where the calculated exchange coupling parameters $J_{ij}$ (the atoms \emph{i} and \emph{j} are labeled according to Fig.~\ref{corral1abcde}(a)) are strongest and antiferromagnetic between nearest neighbors ($J_{12}=-52$ meV), with a smaller long-range interaction that oscillates between ferromagnetic and antiferromagnetic couplings ($J_{24}=0.1$ meV, $J_{13}=0.5$ meV and $J_{34}=-0.04$meV).  

Since the nearest-neighbour exchange coupling is dominant, the ring shaped structure could be seen as an equivalent to a 1d antiferromagnetic Heisenberg chain and thus the collinear ground state is expected. A more interesting situation is when an additional atom is added to the cluster.
 
\begin{figure}[htp]
\begin{center}
\includegraphics*[width=0.7\linewidth]{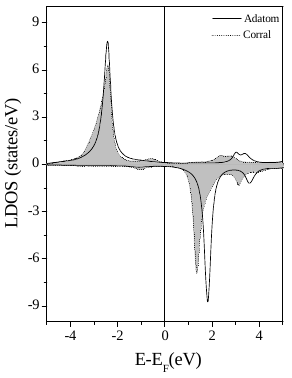}
\caption{\label{ldos-ad}  Local density of states (LDOS) for a Mn adatom on Pt(111) (full line) and for the central Mn atom in the corral displayed in Fig.~\ref{corral1abcde}(b) (gray curve).}
\end{center}
\end{figure}

 For the corral with one extra Mn adatom placed at the center of the cluster, the calculated magnetization profile shown in  Fig.~\ref{corral1abcde}(b) is a consequence of the long-ranged exchange interactions between the central atom and the atoms located in the ring (e.g. $J_{13}=-0.3$ meV and $J_{23}=+0.1$ meV). Note that the magnetization profile in this case is collinear since the extra atom, due to symmetry considerations, does not introduce any frustration to the cluster.
To investigate the influence of the quantum confinement in this nanostructure we compare the
local density of states (LDOS) of the Mn adatom at the center of the corral in Fig.~\ref{corral1abcde}(b) with the LDOS of
an isolated Mn adatom on the Pt(111) surface.  
As shown in Fig.~\ref{ldos-ad}, the peak broadening of the LDOS reflects the electron confinement in the nanostructure, which is due to electron scattering at the border of the nanostructure. The slight modifications in the electronic structure due to this scattering, result in a slightly different magnetic moments which is 4.62$\mu_{B}$/atom for the atom in the middle of the corral and 
 4.69$\mu_{B}$/atom for the isolated adatom.

In Fig.~\ref{corral1abcde}(c) one extra Mn atom is located as nearest neighbor of three Mn atoms in the ring and the magnetization profile is seen to become a noncollinear structure, as a result of the magnetic frustration.\cite{anders} The magnetic moment of the extra atom (labeled 2 in Fig.~\ref{corral1abcde}(c)) is almost perpendicular ($\sim 80^{\circ}$) to the moment of the atom at the edge of the cluster (labeled as 1) and has an angle of $\sim 145^{\circ}$ to the other neighboring atomic moments (atoms labeled 3 and 4). The net moments of these atoms (3 and 4) are parallel and the magnetic configuration of the other atoms in the cluster is noncollinear with a slightly canted antiferromagnetic structure. The exchange interactions between Mn atoms in the cluster are rather strong and antiferromagnetic between nearest neighbors, e.g. for the atoms numbered in Fig.~\ref{corral1abcde}(c) we obtain;
 $J_{12}=-59$ meV, $J_{45}=-57$ meV, and $J_{67}=-52$ meV. One may also note that the exchange interaction is slightly 
stronger for the atoms with more nearest neighbors. A similar magnetic configuration to that of Fig.~\ref{corral1abcde}(c) is obtained for the cluster with two extra atoms inside the corral (Fig.~\ref{corral1abcde}(d)). 
  
Fig.~\ref{corral1abcde}(e) displays the magnetic order, with a noncollinear component, of the Mn corral with one extra atom outside the ring. The angles ($\theta_{ij}$) between the magnetic moment of the Mn extra atom and its two nearest-neighbors are  $\theta_{12}=129^{\circ}$ and $\theta_{24}=135^{\circ}$. The magnetic moments of these neighboring atoms are almost perpendicular ($\theta_{14}=95^{\circ}$) and the remaining atoms of the ring have a canted antiferromagnetic profile, with angles between nearest neighbors around $170^{\circ}$.  By placing the extra atom outside the ring does 
 affect the magnetic ordering of the whole cluster which can be put in contrast with the geometry of an extra atom inside the ring, as in Fig.~\ref{corral1abcde}(c), where the perturbation from a collinear structure was localized to the atoms closest to the additional atom.
 
\par

\begin{figure}[htp]
\begin{center}
\includegraphics*[width=1.0\linewidth]{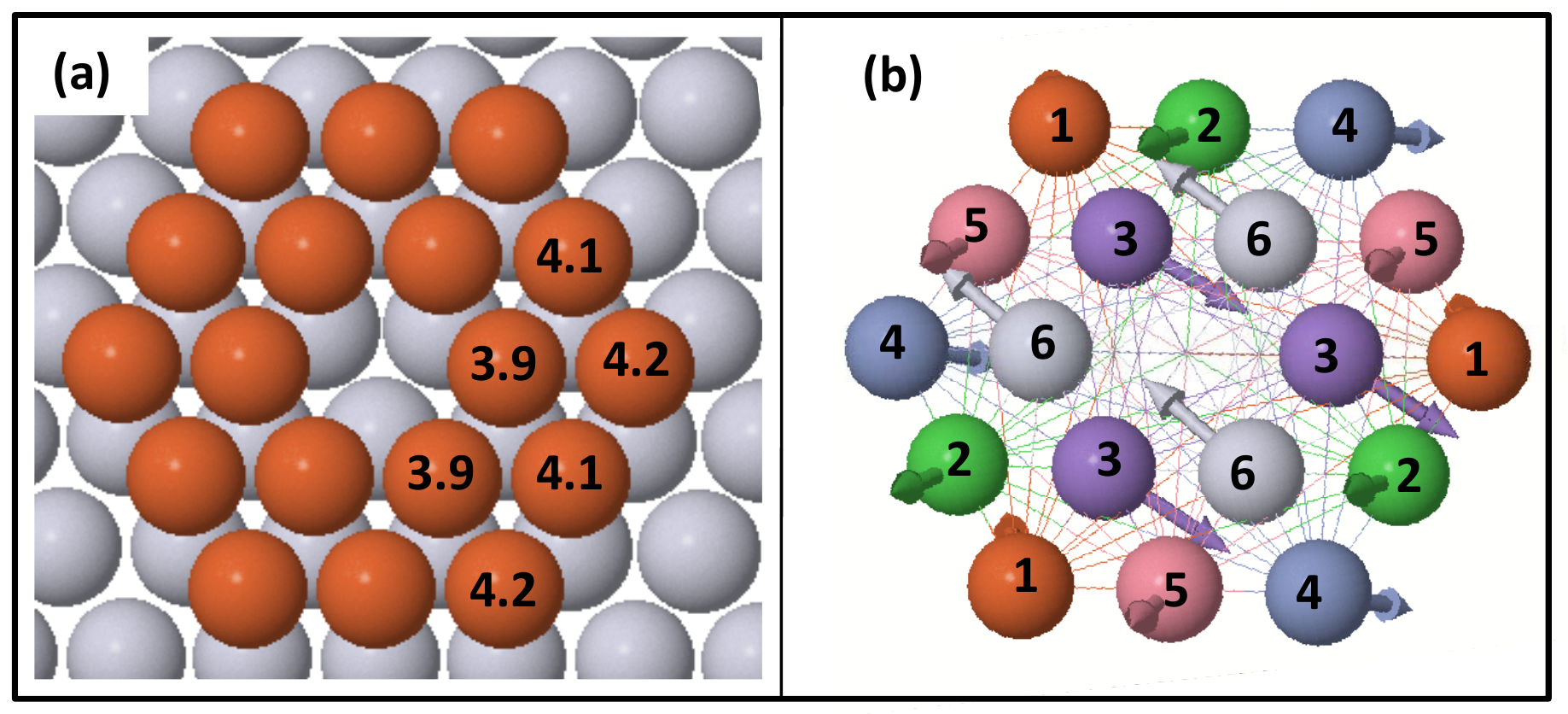}
\caption{\label{doublecorral}(Color online) Mn double-walled corral on a Pt(111) surface. (a) The numbers indicate the
 spin moments (in $\mu_{B}$/atom) of the different atoms. (b) Labeling of the Mn atoms together with the calculated magnetic ordering.}
\end{center}
\end{figure}

 In Fig.~\ref{doublecorral} another interesting scenario is found. Eighteen Mn adatoms forming a double-walled corral were studied. 
As expected, this cluster shows a noncollinear order due to its frustrated geometry. The spin moments for each
non-equivalent Mn site of the structure, for the ferromagnetic 
arrangement, are also
given in Fig.~\ref{doublecorral}(a). The orientations of the six spin sublattices obtained for the atoms 
in the cluster are shown in Table~\ref{table1}, where the atoms are labeled according to Fig.~\ref{doublecorral}(b). The magnetic moments of the outermost atoms are almost perpendicular to the moments of their nearest neighbors located in the inner ring and the remaining atoms of the ring have a canted antiferromagnetic profile, with angles between nearest neighbors around $150^{\circ}$. The angles between nearest neighbors in the inner ring are around $160^{\circ}$. 
 
To investigate the influence of the structural relaxations, for the system shown in Fig.~\ref{doublecorral}, we calculated the electronic structure and magnetic moments for a relaxation of 10\% for the distance between the cluster atoms and the substrate atoms.
The difference of the magnetic moments for the unrelaxed and 10\% relaxed case were found to be less than 2\%. Similarly, for the magnetic configuration we verified that the interaction between the cluster and substrate plays a lesser role compared to the interactions between the cluster atoms. Therefore, for the cluster shown in Fig.~\ref{doublecorral}, the presence of six spin sublattices is maintained even when including a relaxation of 10\%, and the angles between different moments in the cluster are changed with at most a few degrees.
 
\par

\begin{table}[htp]
\caption{The magnetic configuration described by angles between moments for the atoms numbered in the cluster displayed in Fig.~\ref{doublecorral}(b).}

\vspace{0.5cm}
\centering
\label{table1}
\tabcolsep=3pt
\begin{tabular}{c|cccccc}
\hline
\hline
Atom&1  & 2 & 3 & 4 & 5 & 6 \\
\hline
  1 &0  &145&112& 57&153&52 \\
  2 &   & 0 &104&154&10 &96 \\
  3 &   &   & 0 &53 &97 &162 \\
  4 &   &   &   & 0 &148&110 \\
  5 &   &   &   &   & 0 &103\\
  6 &   &   &   &   &   & 0 \\
\hline
\hline

\end{tabular}\\
\end{table}

\begin{figure}[htp]
\begin{center}
\includegraphics*[width=0.9\linewidth]{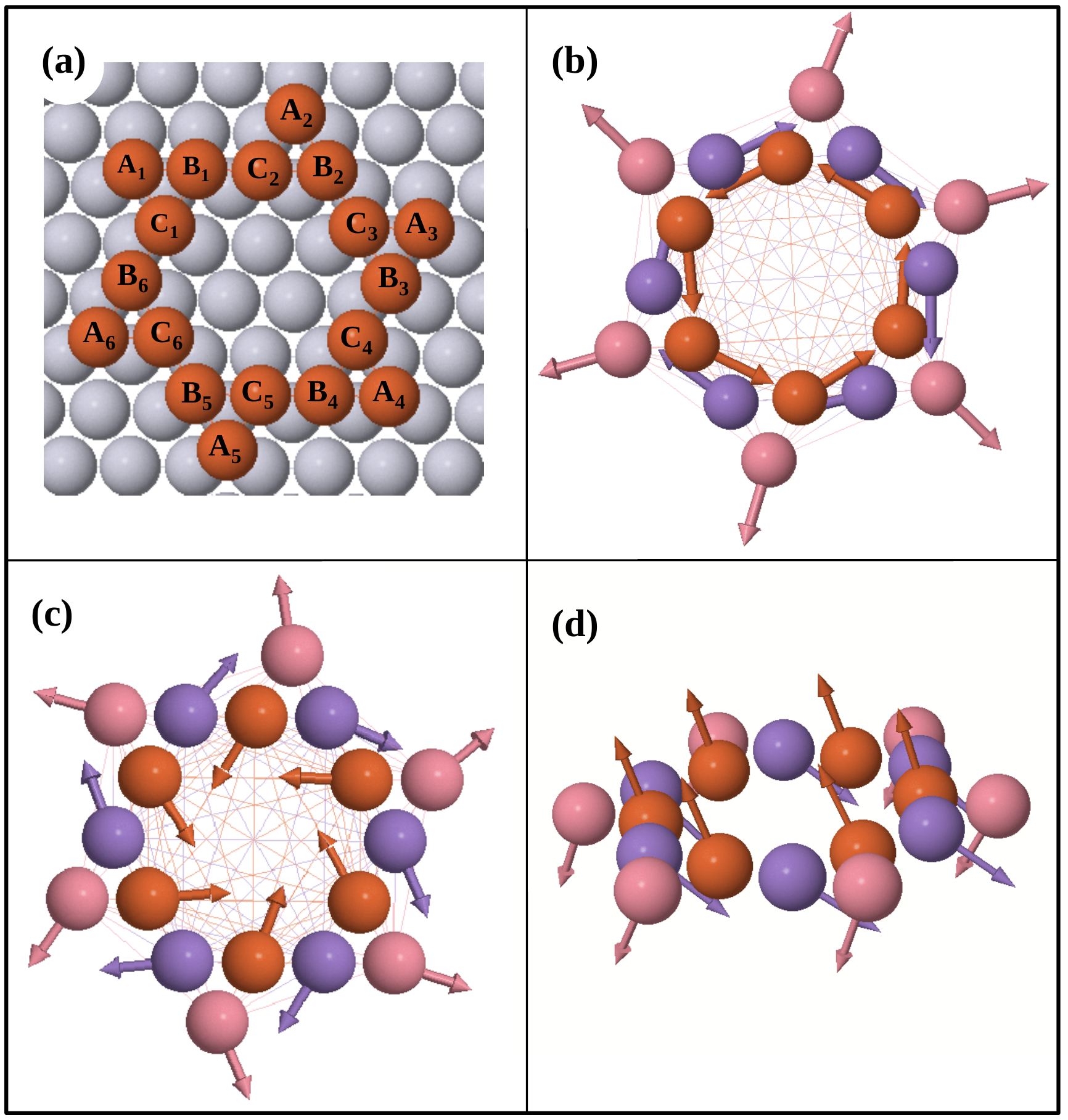}
\caption{\label{kagome}(Color online) Stable magnetic configurations for the Mn corral, composed by six corner sharing triangles ($A_{i}B_{i}C_{i}$), on a Pt(111) surface.}
\end{center}
\end{figure}

Fig.~\ref{kagome}(a) shows the geometry of the Mn corral composed by six corner sharing triangles ($A_{i}B_{i}C_{i}$).
For this geometry we find several local total energy minima which are stable, or metastable, and the differences in energy between these magnetic structures are of the order of a few meV/at. In Figs.~\ref{kagome}(b), (c) and (d) the  magnetic configurations are presented, where the non-planar structure in Fig.\ref{kagome}(c) has the lowest energy.
The differences in energy between these configurations are: $(E_{d}-E_{c})=-4.3meV/atom$, 
$(E_{d}-E_{b})=-2.7meV/atom$ and $(E_{b}-E_{c})=-1.6meV/atom$, where $E_{b}$, $E_{c}$ and $E_{d}$ refer to the energy of configurations shown in Fig.~\ref{kagome}(b), (c) and (d), respectively.
In all cases, the magnetic structures present three sublattices, showed by different colors, such that no
two neighboring sites have the same color. 
Also, for each magnetic structure, the sum of the three angles between the magnetic moments of Mn atoms in each triangle ($A_{i}B_{i}C_{i}$) is always equal to $360^{\circ}$.
For the magnetic configuration shown in Fig.~\ref{kagome}(b), the magnetic moments of the Mn neighboring atoms 
in the 12 atoms ring are almost antiparallel ($\sim 170^{\circ}$) for atoms in  different triangles ($B_{i}$ and $C_{j}$), and $\sim 110^{\circ}$ for atoms in  the same triangles ($B_{i}$ and $C_{i}$). The angles between the magnetic moment of each Mn extra atom outside the ring ($A_{i}$) and its two nearest neighbors ($B_{i}$ and $C_{i}$) are $\sim 120^{\circ}$ and $\sim 130^{\circ}$, respectively.  The magnetic structures presented in Fig.~\ref{kagome}(b) and (c) are almost planar and can be described as a $Z_{2}$ vortex \cite{zhito,japan}, in which the pattern vortex corresponds to a configuration with magnetic moments  performing a $2\pi$ rotation around the vortex center. For the non-planar magnetic configuration displayed in Fig.~\ref{kagome}(d), the angles between magnetic moments of Mn nearest neighbors atoms $B_{i}$ and $C_{j}$ are around $140^{\circ}$,
and the angles between the magnetic moment of each Mn labeled as $A_{i}$ and its two nearest neighbors ($B_{i}$ and $C_{i}$) are $\sim 80^{\circ}$ and $\sim 140^{\circ}$, respectively. 
 
Using the \textit{Jij}'s obtained by first principles for the system shown in Fig.~\ref{kagome}, we performed a Monte Carlo ground state search, obtaining a collinear antiferromagnetic ordering. This indicates that the isotropic exchange interactions \textit{Jij} are not the only important interactions present for this system and are not enough to describe the ground state for this system. Therefore, we expect that the present work can motivate further studies to investigate other interactions that can be significant for such systems.  

\par

%
\section{Conclusions}
In this paper, using the first principle non-collinear RS-LMTO-ASA method, we have presented theoretical predictions of the magnetic configurations of several Mn corrals placed on the Pt(111) surface. We have found that neighboring Mn atoms present strong antiferromagnetic exchange interactions, leading to collinear magnetic configurations when the geometrical frustration is avoided, and novel non-collinear ordering for corrals on a frustrated lattice. In particular, the Mn corral composed by six corner sharing triangles shows a new perception of magnetic structures in nanomagnets, presenting very complex stable magnetic configurations, which can be described by topological structures, as $Z_{2}$ vortices. We expect that the present work can motivate further studies on topological nanomaterials for novel spin devices.

\section{Acknowledgements}
We acknowledge financial support from the Swedish Research
Council, the Swedish Foundation for Strategic Research, FAPESPA, CAPES and CNPq, Brazil. 
O.E. also acknowledges the ERC for support. The calculations were performed at
the computational facilities of the LCCA, University of S\~ao
Paulo, and at the CENAPAD, University of Campinas, SP, Brazil.

\end{document}